\documentclass[prl,aps,preprint,epsfig]{revtex4}
\begin{document}
\title{Magnetic properties of Sr$_{2}$Fe$_{1+x}$Mo$_{1-x}$O$_{6}$ with -1 $\leq$ $x$ $\leq$ 0.25 }

\author{Dinesh Topwal}
\address{Solid State and Structural Chemistry Unit, Indian Institute of Science, Bangalore - 560012, India.}
\author{D. D. Sarma}
\altaffiliation{Also at : Jawaharlal Nehru Centre for Advanced
Scientific Research, Bangalore 560064, India.}
\email{sarma@sscu.iisc.ernet.in}
\address{Solid State and Structural Chemistry Unit, Indian Institute of Science, Bangalore - 560012, India.}

\author{H. Kato$^{1,}$}
\altaffiliation{Present Address: Material Science and Technology
Laboratory, Fuji Electric Advanced Technology, Tokyo 191-8502,
Japan}
\author{Y. Tokura$^{1,2}$}
\address{$^{1}$Correlated Electron Research Center, AIST, Tsukuba 305-8562, Japan.}
\address{$^{2}$Department of Applied Physics, University of Tokyo,Tokyo 113-8656, Japan.}

\author{M. Avignon}
\address{Laboratoire d'Etudes des Proprie´te´s Electroniques des Solides, CNRS, BP 166, 38042 Grenoble Cedex 9, France}

\begin{abstract}

 We have synthesized the solid solution, Sr$_{2}$Fe$_{1+x}$Mo$_{1-x}$O$_{6}$ with
 -1 $\leq$ $x$ $\leq$ 0.25, the composition  $x$~=~0 corresponding to the well-known
 double perovskite system Sr$_{2}$FeMoO$_{6}$. We report structural and
 magnetic properties of the above system, exhibiting systematic variations across the series.
 These results restrict the range of models that can explain magnetism in
 this family of compounds, providing an understanding of the
 magnetic structure.

\end{abstract}


\maketitle

\section{Introduction}

Half metallic ferromagnetic oxides\cite{tokurabook} have attracted
extensive attention not only as a source of fully polarized charge
carriers for spintronics applications, but also as potential
candidates for memory devices by the virtue of their large
magnetoresistance (MR). Recently, a double perovskite,
Sr$_{2}$FeMoO$_{6}$, belonging to this general family of
half-metallic ferromagnetic oxides, has shown spectacularly large
MR even at the room temperature and at relatively small applied
magnetic fields\cite{kobayashiNAT98} compared to the extensively
investigated class of magnetoresistive manganites, due to a
substantially enhanced Curie temperature (T$_{c}$). The enhanced
T$_{c}$ and the origin of ferromagnetism in Sr$_{2}$FeMoO$_{6}$
and related compounds have been explained\cite{sarmaprl00} in
terms of a kinetically driven mechanism. It is interesting to note
that the same mechanism has been later
invoked\cite{kana_terajps01,sarmaCUR.OP01,priya} to explain
ferromagnetism in the recently discovered dilute magnetic
semiconductors (DMS), such as Mn-doped GaAs, thereby connecting
two apparently disparate classes of compounds.

A perfectly ordered lattice of Sr$_{2}$FeMoO$_{6}$ consists of
alternating FeO$_{6}$ and MoO$_{6}$ octahedra along all the three
cubic axes of the perovskite structure. It is believed that the
fully-ordered compound is a half-metallic ferromagnet with nominal
ionic configurations of Fe$^{3+}$ and Mo$^{5+}$. The measured
value of saturation magnetization, $M_{s}$, in normally prepared
samples is invariably
found\cite{kobayashiNAT98,rayjpcon01,tomiokaprb00} to be lower
than the expected value of $4~\mu$$_{B}$ per formula unit (f.u.)
from the half-metallic state. This is due to the inevitable
presence of mis-site disorders, where some Fe and Mo interchange
their crystallographic positions. However, the microscopic origin
of the reduction in $M_{s}$ is still not entirely clear. Both
classical Monte carlo simulations\cite{ogaleAPL99} and quantum
mechanical band structure calculations\cite{T.SahaPRB001} indeed
predicted a reduction of $M_{s}$ as a function of mis-site
disorder, but the underlying reasons for this reduction are very
different in these two proposals. In one case,\cite{ogaleAPL99} it
is assumed that the moment reduction is only due to
antiferromagnetically coupled Fe pairs whenever Fe-O-Fe bonds are
generated by such mis-site disorders. In this view, the conduction
band presumably retains its polarization to a large extent. In
contrast, the band structure approach\cite{T.SahaPRB001}
attributed the reduction of the moment to strong depolarization
effects at each site, though all the Fe sites were found to be
ferromagnetically coupled. The mis-site disorder and its effect on
the conduction band are not only important in the context of
spin-injection devices, the magnetoresistive properties of these
samples also appear to be strongly influenced by these
factors.\cite{ddMR_sfmo,marprl01} Therefore, it is important to
investigate experimentally the underlying details of the magnetic
structure and the effect of mis-site disorder to obtain a
definitive understanding of magnetism in this class of compounds.

There have been some efforts to study samples with different
extents of
disorder;\cite{balcellsAPL2001,jnavarrocon2001,DsanchezJMMM02,sarmaSSC00,ddMR_sfmo}
however, in view of the difficulty to have a microscopic and
detailed control of such a statistical process, we have adopted a
different route to probe issues raised here. We have prepared
Sr$_{2}$Fe$_{1+x}$Mo$_{1-x}$O$_{6}$ over a wide range of
compositions (-1 $\leq$ $x$ $\leq$ 0.25). The ideal $x$ = 0
composition corresponds to the maximal Fe content possible in this
crystal structure, while avoiding a nearest neighbor Fe-O-Fe
arrangement. Therefore, in the Fe deficient regime ($x < $0) we
control the average separation between the magnetic ions by
changing $x$. In the Fe rich composition ($x > $0), we replace
some of the Mo's with Fe in the ordered structure, thereby forcing
the additional Fe ions necessarily to form Fe-O-Fe 180$^{0}$
bonds. Thus, we obtain a control on number of such bonds
introduced in the system by controlling $x$. We report here a
detailed investigation of magnetic properties of this family of
compounds. Our results conclusively show that while Fe-O-Fe bond
is indeed antiferromagnetically coupled, the conduction bands in
these system are also far from being 100\% polarized.
Interestingly, this system shows a rapid increase of $T_{c}$ with
increasing $x$ till approximately $x$ = 0, followed by a near
saturation.

\section{Experiment}

Polycrystalline samples of Sr$_{2}$Fe$_{1+x}$Mo$_{1-x}$O$_{6}$ for
$x > $0 were prepared by melting a stoichiometric mixture of
SrCO$_{3}$, Fe$_{2}$O$_{3}$, MoO$_{3}$ and Mo in an inert gas (Ar)
arc furnace.\cite{sarmaSSC00} The obtained ingots were crushed and
made into pellets. These pellets were sintered at 1250~$^{0}$C in
a reducing atmosphere consisting of 1\% H$_{2}$ and 99\% Ar flow
for 16 hours to achieve the maximum ordering between Fe and Mo and
to remove a small amount ($<$2\%) of SrMo0$_{4}$ present at times.
The samples with lower Fe concentrations ($x$ $\leq$0) were made
by the normal solid-state synthesis.\cite{kobayashiNAT98} In this
case, the stoichiometric mixture of SrCO$_{3}$, Fe$_{2}$O$_{3}$
and MoO$_{3}$ was first heated in air at 900$^{0}$~C for 3 hrs.,
followed by a reaction in the pellet form at 1250$^{0}$~C for 10
hrs. in a mixture of Ar and H$_{2}$ and then furnace-cooled in the
flow of the same reducing gas mixture. The above process of
heating in the reducing atmosphere was repeated 2-3 times with
intermediate grindings to ensure a homogeneous, pure phase. The
ratio of H$_{2}$:Ar varied from 15:85 for $x$ = -1 sample to 2:98
for $x$ = 0 sample. X-ray powder diffraction measurements
established the formation of a single phase material in every
case. We also used energy dispersive analysis of x-rays (EDAX) in
conjunction with secondary electron microscopy (SEM) extensively,
to obtain the concentration ratio between Fe and Mo in different
grains as well as at several points within the same grain for each
sample in order to establish the homogeneity of the samples.
Magnetic properties of the sample were measured using Quantum
Design's Magnetic Properties Measurement System (MPMS)

\section{Results and Discussion}

Fig.~1 shows the powder X-ray diffraction patterns for various
compositions of Sr$_{2}$Fe$_{1+x}$Mo$_{1-x}$O$_{6}$. It is
well-known\cite{sarmaSSC00} that the intense supercell reflection
peaks corresponding to the double perovskite structure appear at
2$\theta$ =~19.4$^{0}$ and 37.8$^{0}$ in Sr$_{2}$FeMoO$_{6}$. Both
these diffraction peaks gradually lose intensity with a deviation
of $x$ from 0, as seen in Fig.~1. The observation of these
diffraction peaks over an extended range of compositions, -0.25
$\leq$ $x$ $\leq$ 0.25 in Sr$_{2}$Fe$_{1+x}$Mo$_{1-x}$O$_{6}$ is
interesting and suggests that Fe preferentially occupies one
sublattice and Mo the other one even for compositions away from
the ideal Fe:Mo::1:1 composition. The x-ray diffraction pattern
establishes a systematic reduction in the unit cell parameters, as
evidensed by a monotonic shift of all diffraction peaks to higher
angles with increasing $x$. We plot the variation of the lattice
parameter, $a$, as obtained from Rietveld refinements as a
function of the composition, $x$, in the inset to Fig. 1. We have
also shown from literature\cite{srfeo3} the corresponding lattice
parameter of SrFeO$_{3}$ which is one end member of this series,
Sr$_{2}$Fe$_{1+x}$Mo$_{1-x}$O$_{6}$, with $x$ = 1. It is
interesting to note that the variation of $a$ over the range of
compositions studied here, namely -1 $\leq$ $x$ $\leq$ 0.25, is
linear with the composition $x$. However, the extrapolation of
this linear trend to the $x$ = 1 end-point is in complete
disagreement with the observed result for SrFeO$_{3}$, suggesting
a drastic or sudden change of valency between $x$ = 0.25 and $x$ =
1 consistent with the fact that Fe is in Fe$^{4+}$ state in
SrFeO$_{3}$ ({\it i.e.} $x$ = 1) and in the Fe$^{3+}$ state in
Sr$_{2}$FeMoO$_{6}$ ({\it i.e.} $x$ = 0). Given the well known
stability of the half-filled orbitals by maximizing the spin
moment, it is reasonable to expect that Fe would prefer to remain
in Fe$^{3+}$ 3$d^{5}$ state, whenever possible. However, a
substitution of Mo$^{5+}$ by Fe$^{3+}$ or {\it vice versa}, as in
Sr$_{2}$Fe$_{1+x}$Mo$_{1-x}$O$_{6}$ with $x\neq0$, cannot satisfy
the charge neutrality without changing the valency of Mo, provided
Fe remains in the trivalent state. Mo is known to adopt readily
valence state between 4+ and 6+. Simple considerations then show
that Fe can remain in the trivalent state for
-1~$\leq$~$x$~$<$~1/3 with Mo continuously changing its valency
from Mo$^{6+}$ for $x$ = 1/3 to Mo$^{4+}$ for $x$ = -1. This
suggests that Fe cannot retain its trivalent state for $x \geq$
1/3, as Mo cannot take up a valency larger than 6+. Thus, it
appears that a systematic and continuous change in the lattice
parameters for -1 $\leq x \leq$ 0.25 is due to the progressive
replacement of Fe by Mo, retaining Fe in the trivalent state,
while the discontinuous change in the lattice parameter for
SrFeO$_{3}$ is due to the change of Fe valency. Recent
spectroscopic studies\cite{XAS} of the solid solution over the
relevant $x$ range indeed support the existence of essentially the
same charge state of Fe over this range of compositions.

The inset to Fig.~2. shows the temperature dependence of
magnetization for an applied field of $\mu_{o}H~=~0.1$~T. It is
evident from these plots that all the samples are ferromagnetic,
exhibiting a strong dependence of the transition
temperature,\cite{note} $T_{c}$ on the composition, $x$. $T_{c}$'s
(Fig.~2) exhibit a nearly linear increase with $x$ between $-0.5$
and $0.05$ and a very slow increase or near saturation for larger
$x$ compositions. We can understand the monotonic increase of
$T_{c}$ with $x$ in the $x\leq 0$ regime in terms of the
increasing gain in kinetic energy concomitant with increasing
spin-polarization of itinerant electrons in the\textit{\
ferromagnetic} Fe background, arising from the increasing
concentration of Fe and a decreasing Fe-Fe separation. Considering
an idealized, fully ordered structure with alternating Fe and Mo
along the three cubic axes, all Fe ions substituting Mo beyond
$x=0$ must necessarily give rise to Fe-O-Fe-O-Fe sequences. Strong
superexchange interactions between Fe-O-Fe bonds should enhance
the ferromagnetic coupling between Fe ions in the same sublattice,
therefore increasing $T_{c}$ as obtained in the case of a small
extent of mis-site disorder.\cite{Allub} On the other hand, the
decreasing number of itinerant electrons, $(1-3x)$, should lower
$T_{c}$. As a result of these two opposing influences, $T_{c}$
shows a near saturation close to $x=0$.

The inset to Fig.~3 shows the field dependence of the
magnetization at 4.2~K. All the magnetization curves are
qualitatively similar, establishing a soft ferromagnetic ground
state for all the compositions. However, there is evidently a
marked variation in the magnetic moment with the composition, $x$.
We have plotted the saturation magnetization per formula unit
(f.u.) at 5~T for all the samples as a function of $x$ in the
mainframe of Fig.~3. The saturation magnetization, $M_{s}$ shows
the remarkable behavior of first increasing nearly linearly with
$x$ up to $x=0$ and then decreasing once again linearly with $x$,
as suggested by the linear fits (solid lines) through the
experimental data (open circle). This dependence, as we next show,
severely restricts the available models to explain the magnetic
structure, thereby providing a detailed understanding of magnetism
and the role of anti-site defects in this family of compounds.

We consider a strongly correlated description for Fe with the
nominal configuration $3d^{5}$ forming high-spin $S=5/2$ localized
spins, and Mo$^{6+}$ cores. The additional $(1-3x)$ nominally
Mo-electrons are itinerant, consistent with all compounds studied
here being metallic. All the five Fe $d$-orbitals of one spin
channel being occupied, itinerant electrons can hop to Fe sites
only with an antiparallel orientation with respect to the
localized spins; thus, the hopping stabilization of itinerant
electrons also leads to its spin-polarization in the idealized,
fully ordered
Sr$_{2}$FeMoO$_{6}$.\cite{sarmaprl00,Allub,kana_terajps01,sarmaCUR.OP01}
It is also clear from this discussion that the polarization of
each Mo site depends on the immediate Fe environment and its
localized spin structure in the real system with anti-site
disorders.\cite{Allub} In the $x=0$ fully ordered case, Fe-Mo
hopping gives rise to the minority band and half-metallic ground
state.\cite{sarmaprl00,kana_terajps01,sarmaCUR.OP01} A
\emph{qualitative} understanding of the observed variation of the
magnetic moment can be obtained easily. For $x\leq 0$ regime, the
saturation magnetization $M_{s}$ increases with $x$ primarily
because the number of local moments increases, and all fe moments
are ferromagnetically coupled. In the regime $x\geq 0$, as the Fe
content increases, we find a rapid decrease of $M_{s}$. This is
readily understood in terms of the additional Fe being necessarily
connected to the next Fe by $180^{0}$ Fe-O-Fe bonds and
consequently, being antiferromagnetically coupled. However, a more
detailed and quantitative understanding requires separating two
contributions to the $M_{s}$, namely those ($M_{Fe}$) from
localized moments of Fe and those ($M_{Mo}$) from conduction
electrons. The local moment contribution,
$M_{Fe}=5(n_{Fe}^{1}-n_{Fe}^{2})$ with $n_{Fe}^{1}$ and
$n_{Fe}^{2}$ being the proportion of Fe on sublattice $1$ and $2$,
belonging to Fe and Mo, respectively in the fully ordered
Sr$_{2}$FeMoO$_{6}$. We define the order parameter, $a$, as the
probability to find the minority component in its proper position.
Therefore, the number of Fe in the right position for $x\leq0$,
$n_{Fe}^{1}=a(1+x)$, consequently, $n_{Fe}^{2}=(1-a)(1+x)$, as
well as $n_{Mo}^{1}=1-a(1+x)$ and $n_{Mo}^{2}=a(1+x)-x$. Similarly
in the case $x\geq 0$ the corresponding number for Mo is
$n_{Mo}^{2}=a(1-x)$ and $n_{Mo}^{1}=(1-a)(1-x)$, with
$n_{Fe}^{2}=1-a(1-x)$ and $n_{Fe}^{1}=x+a(1-x)$. Therefore, we get
$M_{Fe}=5(2a-1)(1+x)$ for $x\leq 0$ and $5(2a-1)(1-x)$ for $x\geq
0$. We have determined the degree of order, $a$, in all our
samples using Rietveld refinement as 0.65, 0.83, 0.92, 0.89, 0.91,
0.90, 0.86 and 0.83 for $x=-0.5, -0.25, -0.1, -0.05, 0.0, 0.05,
0.1$ and $0.25$, respectively, completely determining the local
moment contribution to the total moment. In order to estimate now
the contribution of the conduction electron to $M_{s}$, we first
note that the number of conduction electrons per Mo will vary as
$n=(1-3x)/(1-x)$ in Sr$_{2}$Fe$_{1+x}$Mo$_{1-x}$O$_{6}$, in view
of Fe retaining its localized trivalent state. In order to
estimate the polarization $M_{Mo}$ of the \emph{n} conduction
electrons, we first note that Mo surrounded by 6 Fe's as in
ordered Sr$_{2}$FeMoO$_{6}$, \emph{i.e.} $a=1$ and $x=0$, yields a
fully polarized conduction band, leading to $M_{Mo}=-1~\mu _{B}$
per f.u. The presence of Mo in the Fe sublattice, either due to
disorder or due to Mo excess in the $x\leq 0$ regime, opens up
Mo$^{1}$ -Mo$^{2}$ hopping channels \emph{via} oxygen states, as
in the end-member SrMoO$_{3}$, thereby depolarizing the conduction
band. Simple Green functions calculations in presence of
disorder\cite{Allub2} support this view. In the limit of
SrMoO$_{3}$ ($x=-1$) with Mo surrounded only by Mo, the conduction
moment is zero. Interpolating linearly between these two limits,
by assuming the depolarization of the Mo electrons to be
proportional to the number of Mo sites, in the other sublattice,
we can write
$M_{Mo}=-[n_{Mo}^{1}+n_{Mo}^{2}-2n_{Mo}^{1}n_{Mo}^{2}](1-3x)/(1-x)$,
with the overall negative sign on $M_{Mo}$ representing the
antiferromagnetic coupling between $M_{Fe}$ and $M_{Mo}$. The
total magnetic moment per formula unit, $M_{s}$=$M_{Fe}$+$M_{Mo}$,
is then readily evaluated as a function of only the composition,
$x$, along with experimentally determined values of a, the order
parameter. We have plotted these estimated $M_{s}(x)$ with closed
triangles in the main frame of Fig.~3. Considering that the
expression for $M_{s}$ is fixed by $x$, with no adjustable
parameter in the model, the agreement between the experimental and
the model results is exceptionally good. In the same figure, we
have also shown the variation in the moment, $M_{Mo}$ arising from
the conduction band (star) with the composition $x$. Considering
that the conduction states are only nominally Mo states and in
reality, these have nearly equal contributions from Mo, Fe and
O,\cite{T.SahaPRB001} we anticipate a moment of about one-third of
$M_{Mo}$ to be associated with the Mo sites, leading to about
-0.28~$\mu _{B}$ for the Mo sites in the $x = 0$ compound in the
present case. This is consistent with the estimates of Mo moment
in Sr$_{2}$FeMoO$_{6}$.\cite{xmcdprl,besse}

\section{Conclusion}

We have synthesized the solid solution,
Sr$_{2}$Fe$_{1+x}$Mo$_{1-x}$O$_{6}$, over a wide range of
compositions. X-ray diffraction results establish a trivalent
Fe$^{3+}$ state over the entire range of $x$ with the charge
neutrality being maintained by a continuous changing of Mo
valency. An analysis of the magnetic moment as a function of $x$
supports a very specific model, where any Fe at the "wrong"
crystallographic site is coupled anti-parallel to the Fe moments
at the correct site. Additionally, Mo is found to depolarize to an
extent proportional to the number of Mo sites in the near-neighbor
coordination shell. These results resolve the conflicting views
proposed earlier concerning the magnetic structure of such double
perovskite oxides.

\section{Acknowledgments}

This project is supported by the Department of Science and
Technology and Board of Research in Nuclear Sciences, Government
of India.

\section{Figure captions}

Fig.~1 X-ray powder diffraction patterns of
Sr$_2$Fe$_{1+x}$Mo$_{1-x}$O$_6$. Inset shows variation of lattice
parameter, $a$, as a function of the composition. The data for
$x=1$ corresponding to SrFeO$_{3}$ is taken from reference 17.

Fig.~2 Plot of $T_c$ \emph{vs} $x$. The dotted line is a guide to
the eye. The inset shows the temperature dependence of
magnetization for an applied field of $\mu$$_{o}H$ = 0.1 T, for
various compositions.

Fig.~3  Experimental saturation magnetization (open circle) at
4.2~K and 5~T as a function of the composition, $x$, compared with
the calculated ones (solid triangle) based on the model presented
in the text. Stars represent the calculated moment contribution
from the conduction electrons per formula unit. The inset shows
$M(H)$ at 4.2~K for various $x$.


\begin{thebibliography}{99}
\bibitem{XAS} Unpublished x-ray absorption spectra at the
Fe~$L_{2,3}$ edge and M\"{o}sbauer spectra of these samples from our
group clearly support this view.
\bibitem{note} We estimate $T_{c}$ by the peak in the derivative curve, $dM/dT$;
the usual definition of $T_{c}$ by the temperature where $M(T)$
begins to rise rapidly would yield transition temperatures that
are typically higher than the present estimates. The two lowest
$x$ compositions suggest the presence of low concentrations of
magnetically impure phases by the smooth distribution of finite
$M$ at higher temperatures, though the main compositions clearly
show sharp variations in $M(T)$ defining $T_{c}$.
\end{thebibliography}
\end{document}